\documentclass{article}
\usepackage{amsmath}
\usepackage{graphicx}
\usepackage{amsfonts}
\usepackage{amssymb}
\newtheorem{theorem}{Theorem}

\newtheorem{lemma}[theorem]{Lemma}

\begin{document}

\title{Exact form factors for the scaling $Z_{N}$-Ising\\and the affine $A_{N-1}$-Toda quantum field theories}
\author{H. Babujian \thanks{ Permanent address: Yerevan Physics Institute, Alikhanian
Brothers 2, Yerevan, 375036 Armenia.} \thanks{ e-mail: babujian@lx2.yerphi.am,
babujian@physik.fu-berlin.de}~ and M. Karowski \thanks{ e-mail: karowski@physik.fu-berlin.de}\\Institut f\"ur Theoretische Physik\\Freie Universit\"at Berlin,\\Arnimallee 14, 14195 Berlin, Germany}
\maketitle
\begin{abstract}
Previous results on form factors for the scaling Ising and the sinh-Gordon
models are extended to general $Z_{N}$-Ising and affine $A_{N-1}$-Toda quantum
field theories. In particular result for order, disorder parameters and
para-fermi fields $\sigma_{Q}(x),\;\mu_{\tilde{Q}}(x)$ and $\psi_{Q}(x)$ are
presented for the $Z_{N}$-model. For the $A_{N-1}$-Toda model form factors for
exponentials of the Toda fields are proposed. The quantum field equation of
motion is proved and the mass and wave function renormalization are calculated
exactly.\\[8pt]PACS: 11.10.-z; 11.10.Kk; 11.55.Ds\newline Keywords: Integrable
quantum field theory, Form factors
\end{abstract}

In the framework of the bootstrap program the quantum field theories are
defined by the S-matrices. However, we may motivate the models as follows: The
$Z(N)$-Ising quantum field theory in 1+1 dimensions is considered as the
scaling limit of a classical statistical lattice model in 2-dimensions given
by the partition function
\[
Z=\sum_{\left\{  \sigma\right\}  }\exp\left(  -\frac{1}{kT}\sum_{\left\langle
ij\right\rangle }E(\sigma_{i},\sigma_{j})\right)  \,;\;\sigma_{i}\in\left\{
1,\omega,\dots,\omega^{N-1}\right\}  ,\;\omega=e^{2\pi i/N}%
\]
as a generalization of the Ising model. It was conjectured by K\"{o}berle and
Swieca \cite{KS} that there exists a $Z(N)$-invariant interaction
$E(\sigma_{i},\sigma_{j})$ such that the resulting quantum field theory is
integrable. This model has also been discussed as a deformation \cite{Za,Fa}
of a conformal $Z_{N}$ para-fermi field theory \cite{FZ}. The classical
$A_{N-1}$-Toda model is defined by the Lagrangian
\[
\mathcal{L}=\frac{1}{2}\left(  \partial_{\mu}\vec{\varphi}\right)  ^{2}%
+\frac{\alpha}{\beta^{2}}\sum_{j=0}^{N-1}\exp\left(  \beta\vec{\alpha}_{j}%
\vec{\varphi}\right)
\]
where $\vec{\varphi}=\left(  \varphi^{1},\dots,\varphi^{N-1}\right)  $ are
real fields, $\vec{\alpha}_{j}\;(|\vec{\alpha}_{j}|=\sqrt{2})$ are the simple
positive $A_{N-1}$-roots and $\vec{\alpha}_{0}=-\sum_{j=1}^{N-1}\vec{\alpha
}_{j}$. The field equations are
\begin{equation}
\square\vec{\alpha}_{j}\vec{\varphi}+\frac{\alpha}{\beta}\left(  2e^{\beta
\vec{\alpha}_{j}\vec{\varphi}}-e^{\beta\vec{\alpha}_{j+1}\vec{\varphi}%
}-e^{\beta\vec{\alpha}_{j-1}\vec{\varphi}}\right)  =0\,. \label{fe}%
\end{equation}

The $Z_{N}$-Ising model in the scaling limit and the affine $A_{N-1}$-Toda
model posses the same particle content. There are $N-1$ types of particles:
$a=1,\dots,N-1$ of charge $a$, mass $m_{a}=M\sin\pi\frac{a}{N}$ and $\bar
{a}=N-a$ is the antiparticle of $a$. In particular for $N=2$ the scaling
$Z_{2}$-Ising model is the well investigated model \cite{B-J,BKW} which is
equivalent to a massive free Dirac field theory. The affine $A_{1}$-Toda model
is the sinh-Gordon model which is equivalent to the lowest breather sector of
the sine-Gordon model for imaginary couplings. The n-particle S-matrices
factorize in terms of two-particle ones since the models are integrable. The
two-particle S-matrix for the $Z_{N}$-Ising model has been proposed by
K\"{o}berle and Swieca \cite{KS}. The scattering of two particles of type $1$
is%
\begin{equation}
S^{Z}(\theta)=\frac{\sinh\frac{1}{2}(\theta+\frac{2\pi i}{N})}{\sinh\frac
{1}{2}(\theta-\frac{2\pi i}{N})}\,. \label{1}%
\end{equation}
The two-particle S-matrix for the $A_{N-1}$-Toda model has been proposed by
Arinshtein, Fateev and Zamolodchikov \cite{AFZ} (see also \cite{BCDS}). The
scattering of two particles of type $1$ is
\begin{equation}
S^{T}(\theta)=S^{Z}(\theta)\frac{\sinh\frac{1}{2}(\theta-\frac{\pi i}{N}%
B)}{\sinh\frac{1}{2}(\theta+\frac{\pi i}{N}B)}\frac{\sinh\frac{1}{2}%
(\theta-\frac{\pi i}{N}(2-B))}{\sinh\frac{1}{2}(\theta+\frac{\pi i}{N}%
(2-B))}\,,\;B=\frac{2\beta^{2}}{4\pi+\beta^{2}}. \label{2}%
\end{equation}

Both S-matrices are consistent with the picture that the bound state of $N-1$
particles of type $1$ is the anti-particle of $1$. This will be essential also
for the construction of form factors below.

The form factor bootstrap program has been applied in \cite{BKW} to the
$Z_{2}$-model. Form factors for the $Z_{3}$-model were investigated by one of
the present authors in \cite{K}. There the form factors of the order parameter
$\sigma_{1}$ were proposed up to four particles. Kirilov and Smirnov
\cite{KiSm} proposed all form factors of the $Z_{3}$-model in terms of
determinants. For general $N$ form factors for chargeless states (n particles
of type 1 and n particles of type $N-1$) were calculated in \cite{JKOPS}. Low
particle number form factors of $A_{N-1}$-Toda models\footnote{For other Toda
models see \cite{DM,AMV} and for sinh-Gordon also \cite{FMS,KM}.} where
investigated by Destri and De Vega \cite{DdV}, Oota \cite{Oo} and Lukyanov
\cite{Lu}. In the present letter we present integral representations for all
matrix elements of field operators for the $Z_{N}$-Ising and the $A_{N-1}%
$-Toda models.

For the $Z_{N}$-model we consider the fields $\psi_{Q\tilde{Q}}%
(x)\,,\;(Q,\tilde{Q}=0,\dots,N-1)$ with charge $Q$, spin $Q\tilde{Q}/N$ and
statistics factor\footnote{Another model with nontrivial statistics is the
Federbusch model (see e.g. \cite{CF}).} (with respect to the particle $a=1$)
$e^{2\pi i\tilde{Q}/N}$. There are in particular the order parameters
$\sigma_{Q}(x)=\psi_{Q0}(x)$, the disorder parameters $\mu_{\tilde{Q}}%
(x)=\psi_{0\tilde{Q}}(x)$ and the para-fermi fields $\psi_{Q}(x)=\psi_{QQ}(x)$
(for $Q=1,\dots,N-1$). They satisfy the space like commutation rules:
\begin{align}
\sigma_{Q}(x)\sigma_{Q^{\prime}}(y)  &  =\sigma_{Q^{\prime}}(y)\sigma
_{Q}(x)\nonumber\\
\mu_{\tilde{Q}}(x)\mu_{\tilde{Q}^{\prime}}(y)  &  =\mu_{\tilde{Q}^{\prime}%
}(y)\mu_{\tilde{Q}}(x)\nonumber\\
\sigma_{Q}(x)\mu_{\tilde{Q}}(y)  &  =\mu_{\tilde{Q}}(y)\sigma_{Q}%
(x)e^{\theta(x^{1}-y^{1})2\pi iQ\tilde{Q}/N}\label{cr}\\
\psi_{Q}(x)\psi_{Q}(y)  &  =\psi_{Q}(y)\psi_{Q}(x)e^{\epsilon(x^{1}-y^{1})2\pi
iQ^{2}/N}\,.\nonumber
\end{align}
For the $A_{N-1}$-Toda models we present integral representations for all
matrix elements of normal ordered exponentials of the fields
\[
:\exp\left(  \gamma_{1}\varphi^{1}+\dots+\gamma_{N-1}\varphi^{N-1}\right)  :
\]
where the $\varphi^{a}$ are the fundamental Toda fields.

The generalized form factors $\mathcal{O}_{n}(\theta_{1},\dots,\theta_{n})$
are defined by the vacuum - $n$-particle matrix elements
\[
\langle\,0\,|\,\mathcal{O}(x)\,|\,p_{1},\dots,p_{n}\,\rangle_{a_{1}\dots
a_{n}}^{in}=e^{-ix(p_{1}+\dots+p_{n})}\,\mathcal{O}_{a_{1}\dots a_{n}}%
(\theta_{1},\dots,\theta_{n})
\]
where the $a_{i}$ denote the type (charge) and the $\theta_{i}$ are the
rapidities of the particles $\left(  p_{i}=m(\cosh\theta_{i},\sinh\theta
_{i})\right)  $. This definition is meant for $\theta_{1}>\dots>\theta_{n}$,
in the other sectors of the variables the functions $\,\mathcal{O}_{a_{1}\dots
a_{n}}(\theta_{1},\dots,\theta_{n})$ are given by analytic continuation with
respect to the $\theta_{i}$. General matrix elements are obtained from
$\mathcal{O}_{\underline{a}}(\underline{\theta})$ by crossing which means in
particular the analytic continuation $\theta_{i}\rightarrow\theta_{i}\pm i\pi
$. The form factor equations which have to be solved are\footnote{These
formulae have been proposed in \cite{Sm} as a generalization of formulae in
\cite{KW} and they has been proven in \cite{BFKZ} using LSZ assumptions.}:

\begin{itemize}
\item [(o)]The form factor function $\mathcal{O}_{\underline{a}}%
({\underline{\theta}})$ is meromorphic with respect to all variables
$\theta_{1},\dots,\theta_{n}$.

\item[(i)] It satisfies Watson's equations
\[
\mathcal{O}_{\dots a_{i}a_{j}\dots}(\dots,\theta_{i},\theta_{j},\dots
)=\mathcal{O}_{\dots a_{j}a_{i}\dots}(\dots,\theta_{j},\theta_{i}%
,\dots)\,S_{a_{i}a_{j}}(\theta_{ij}).
\]

\item[(ii)] The crossing relation means for the connected part (see e.g.
\cite{BK}) of the matrix element
\begin{align*}
_{\bar{a}_{1}}\langle\,p_{1}\,|\,\mathcal{O}(0)\,|\,p_{2},\dots,p_{n}%
\,\rangle_{a_{2}\dots a_{n}}^{in,conn.}  &  =\sigma_{\mathcal{O}%
1}\,\mathcal{O}_{a_{1}a_{2}\dots a_{n}}(\theta_{1}+i\pi,\theta_{2}%
,\dots,\theta_{n})\\
&  =\mathcal{O}_{a_{2}\dots a_{n}a_{1}}(\theta_{2},\dots,\theta_{n},\theta
_{1}-i\pi)
\end{align*}
where $\sigma_{\mathcal{O}1}$ is the statistics factor of the operator
$\mathcal{O}$ with respect to the particle 1.

\item[(iii)] The function $\mathcal{O}_{n}({\underline{\theta}})$ has poles
determined by one-particle states in each sub-channel. In particular, if $1$
is the antiparticle of $2$, it has the so-called annihilation pole at
$\theta_{12}=i\pi$ which implies the recursion formula
\[
\operatorname*{Res}_{\theta_{12}=i\pi}\mathcal{O}_{n}(\theta_{1},\dots
,\theta_{n})=2i\,\mathcal{O}_{n-2}(\theta_{3},\dots,\theta_{n})\left(
\mathbf{1}-\sigma_{\mathcal{O}1}S(\theta_{2n})\dots S(\theta_{23})\right)
\,.
\]

\item[(iv)] Bound state form factors yield another recursion formula
\[
\operatorname*{Res}_{\theta_{12}=iu}\mathcal{O}_{ab\dots}(\underline{\theta
})=\sqrt{2}\mathcal{O}_{c\dots}(\theta_{c},\underline{\theta}^{\prime
})\,\Gamma_{ab}^{c}%
\]
where $\Gamma_{ab}^{c}$ is the bound state intertwiner (see e.g. \cite{BK})
defined by
\[
i\operatorname*{Res}_{\theta_{12}=iu}S_{ab}(\theta)=\Gamma_{c}^{ba}\Gamma
_{ab}^{c}%
\]
if $iu$ is the position of the bound state pole.

\item[(v)] Since we are dealing with relativistic quantum field theories
Lorentz covariance in the form
\[
\mathcal{O}_{n}(\theta_{1},\dots,\theta_{n})=e^{s\mu}\,\mathcal{O}_{n}%
(\theta_{1}+\mu,\dots,\theta_{n}+\mu)
\]
holds if the local operator transforms as $\mathcal{O}\rightarrow e^{s\mu
}\mathcal{O}$ where $s$ is the ``spin'' of $\mathcal{O}$.
\end{itemize}

We investigate generalized form factors of an operator $\mathcal{O}(x)$ and
$n$ particles of type $1$ and for simplicity we write $\mathcal{O}%
_{n}(\underline{\theta})=\mathcal{O}_{1\dots1}(\underline{\theta})$. Note that
all further matrix elements with different particle states of the field
operator $\mathcal{O}(x)$ are obtained by the crossing formula (ii) and the
bound state formula (iv). The form factors $\mathcal{O}_{n}(\underline{\theta
})$ are of the form \cite{KW}
\begin{equation}
\mathcal{O}_{n}(\underline{\theta})=K_{n}^{\mathcal{O}}(\underline{\theta
})\prod_{1\leq i<j\leq n}F(\theta_{ij})\,,\quad\left(  \theta_{ij}=\theta
_{i}-\theta_{j}\right)  \label{4}%
\end{equation}
where $F(\theta)$ is the 'minimal' form factor function. It is the solution of
Watsons equation \cite{Wa} and the crossing relation for $n=2$
\begin{equation}%
\begin{array}
[c]{c}%
F(\theta)=F(-\theta)S(\theta)\\
F(i\pi-\theta)=F(i\pi+\theta)
\end{array}
\label{3}%
\end{equation}
with no poles and zeros in the physical strip $0<\operatorname{Im}\theta
\leq\pi$ (and a simple zero at $\theta=0$). We obtain the solutions (see
\cite{KW} for the procedure to solve (\ref{3}))
\begin{align}
F^{Z}(\theta) &  =c_{Z}\exp\int_{0}^{\infty}\frac{dt}{t}\,\frac{2\sinh
t\frac{N-1}{N}\cosh t\frac{1}{N}}{\sinh^{2}t}\left(  1-\cosh t\left(
1-\frac{\theta}{i\pi}\right)  \right)  \label{fz}\\
F^{T}(\theta) &  =\exp\int_{0}^{\infty}\frac{dt}{t}\,\frac{-4\sinh t\frac
{N-1}{N}\sinh t\frac{B}{2N}\sinh t\frac{2-B}{2N}}{\sinh^{2}t}\cosh t\left(
1-\frac{\theta}{i\pi}\right)  \label{ft}%
\end{align}
for the $Z_{N}$ and $A_{N-1}$ S-matrices given by (\ref{1}) and (\ref{2}),
respectively. The normalization constant $c_{Z}$ will be fixed below and
$F^{T}$ is normalized by $F^{T}(\infty)=1$. In eq.~(\ref{4}) $K_{n}%
^{\mathcal{O}}(\underline{\theta})$ is a rational function of the
$e^{\theta_{i}}$ and has the 'physical poles'\footnote{For bound state form
factors there are also higher order 'physical poles' (see e.g.
\cite{CT,BCDS,DM,AMV}).} in $0<\operatorname{Im}\theta_{ij}\leq\pi$
corresponding to the form factor properties (iii) and (iv). The form factor
equations (i) and (ii) hold because of the equations (\ref{3}). We propose
$K_{n}^{\mathcal{O}}(\underline{\theta})$ as linear combinations of the
integrals
\begin{align}
I_{n\underline{m}}(\underline{\theta},p_{n}^{\mathcal{O}}) &  =\frac{1}%
{m_{1}!\dots m_{N-1}!}\left(  \prod_{k=1}^{N-1}\prod_{j=1}^{m_{k}}%
\int_{\mathcal{C}_{\underline{\theta}}}\frac{dz_{kj}}{R}\right)  \label{5}\\
&  \times\prod_{k=1}^{N-1}\left(  \prod_{j=1}^{m_{k}}\prod_{i=1}^{n}%
\phi(z_{kj}-\theta_{i})\prod_{1\leq i<j\leq m_{k}}\tau(z_{ki}-z_{kj})\right)
\nonumber\\
&  \times\prod_{1\leq k<l\leq N-1}\prod_{i=1}^{m_{k}}\prod_{j=1}^{m_{l}}%
\kappa(z_{ki}-z_{lj})p_{n\underline{m}}^{\mathcal{O}}(\underline{\theta
},\underline{z})\nonumber
\end{align}
with
\[
\kappa(z)=\frac{1}{\phi(-z)}\,,\;\tau(z)=\frac{1}{\phi(z)\phi(-z)}\,,\;R=2\pi
i\operatorname*{Res}_{z=0}\phi(z)\,.
\]
The integration contour $\mathcal{C}_{\underline{\theta}}$ encloses the points
$z=\theta_{i}$. The p-functions $p_{n\underline{m}}^{\mathcal{O}}%
(\underline{\theta},\underline{z})$ are symmetric with respect to the
$\theta_{i}$ and the $z_{kj}$ for fixed $j$. They depend on the operator
$\mathcal{O}(x)$ whereas the other functions are determined by the S-matrix
only. For both models the bound state of $N-1$ particles of type $1$ is the
anti-particle of $1$. Together with the form factor recursion relations (iii)
and (iv) this property implies the following equation for the function
$\phi(z)$%
\begin{equation}
\prod_{k=1}^{N-1}\phi\left(  \theta+2\pi ik/N\right)  \prod_{k=0}%
^{N-1}F\left(  \theta+2\pi ik/N\right)  =1\label{6}%
\end{equation}
where $F(\theta)$ is the 'minimal' form factor function, here given by
eqs.~(\ref{fz},\ref{ft}) for the $Z_{N}$- and the $A_{N-1}$-model,
respectively. One easily verifies that the functions
\begin{align*}
\phi^{Z}(z) &  =\frac{1}{\sinh\frac{1}{2}z\sinh\frac{1}{2}(z-2\pi i/N)}\\
\phi^{T}(z) &  =\phi^{Z}(z)\sinh\tfrac{1}{2}\left(  z-i\pi B/N\right)
\sinh\tfrac{1}{2}\left(  z-i\pi\left(  2-B\right)  /N\right)
\end{align*}
solve the equation (\ref{6}) for the $Z_{N}$- and the $A_{N-1}$-model,
respectively, and moreover they satisfy the 'Jost-function property'
\[
\frac{\phi(\theta)}{\phi(-\theta)}=S(\theta)
\]
Equation~(\ref{6}) also determines the normalization constant $c_{Z}$. The
following lemma will be used to construct solutions of the form factor
equations (i) -- (v).

\begin{lemma}
The functions $I_{n\underline{m}}(\underline{\theta},p_{n\underline{m}%
}^{\mathcal{O}})$ satisfy the recursion relation
\begin{multline}
\operatorname*{Res}_{\theta_{N-1N}=iu}\cdots\operatorname*{Res}_{\theta
_{12}=iu}I_{n\underline{m}}(\underline{\theta},p_{n\underline{m}}%
^{\mathcal{O}})=c\left(  \prod_{j=N+1}^{n}\prod_{i=1}^{N}F_{11}^{\min}%
(\theta_{ij})\right)  ^{-1}\label{l}\\
\times I_{n-N\underline{m}-1}(\underline{\theta}^{\prime},p_{n-N\underline
{m}-1}^{\mathcal{O}})\left(  1-\sigma_{\mathcal{O}}\prod_{i=N+1}^{n}%
S(\theta_{Ni})\right)
\end{multline}
if eq.~(\ref{6}) holds, the p-functions are analytic and satisfy for
$\theta_{12}=\dots=\theta_{N-1N}=iu=2\pi i/N$
\begin{align}
p_{n\underline{m}}^{\mathcal{O}}(\underline{\theta},\underline{z}^{\prime})
&  =c^{\prime}p_{n-N\underline{m}-1}^{\mathcal{O}}(\underline{\theta}^{\prime
},\underline{z}^{\prime\prime\prime})\label{7}\\
p_{n\underline{m}}^{\mathcal{O}}(\underline{\theta},\underline{z}%
^{\prime\prime})  &  =\sigma_{\mathcal{O}}\,p_{n\underline{m}}^{\mathcal{O}%
}(\underline{\theta},\underline{z}^{\prime}) \label{8}%
\end{align}
where $c$ and $c^{\prime}$ are a normalization constants, $\sigma
_{\mathcal{O}}$ is the statistics factor of the operator $\mathcal{O}$ with
respect to the particle of type 1 and
\begin{align*}
\underline{z}^{\prime}  &  =\underline{z}^{\prime\prime}=\underline
{z}\,;\;\text{except }z_{i1}^{\prime}=\theta_{i}\,,\;z_{11}^{\prime\prime
}=\theta_{N}\,,\;z_{i+11}^{\prime\prime}=\theta_{i}\\
\underline{z}^{\prime\prime\prime}  &  =\left(  z_{12},\ldots,z_{1m_{1}%
};\ldots;z_{N-12},\ldots,z_{N-1m_{N-1}}\right) \\
\underline{\theta}^{\prime}  &  =\left(  \theta_{N+1},\theta_{N+2}%
,\ldots,\theta_{n}\right)  \,.
\end{align*}
\end{lemma}

The proof of this lemma will be published elsewhere \cite{BK3}. We propose the
following solutions of (i) -- (v) and identify the operators by means of the
quantum numbers as charge, spin and statistics factor. For the Toda model we
also consider the asymptotic behavior and field equations.

\paragraph{The scaling $Z_{N}$-Ising model:}

The scaling $Z_{2}$-Ising model is well known \cite{B-J,BKW}, it is equivalent
to a free fermi field theory. As a generalization we propose for the general
$Z_{N}$-model the form factors for $n$ particles of type $1$ (up to
normalizations) as
\[
\mathcal{O}_{n}(\underline{\theta})=I_{n\underline{m}}^{Z}(\underline{\theta
})\prod_{1\leq i<j\leq n}F^{Z}(\theta_{ij})
\]
where $I_{n\underline{m}}^{Z}(\underline{\theta})$ is defined by eq.~(\ref{5})
in terms of $\phi^{Z}$. The p-functions and $\underline{m}$ are given by the
following correspondences to operators. For $n=Nm+Q$ particles of type $1$ and
$Q,\tilde{Q}=0,\dots,N-1$ the p-functions
\begin{equation}
p_{n\underline{m}}^{Q\tilde{Q}}(\underline{\theta},\underline{z})=\exp\left(
\frac{\tilde{Q}}{N}\sum_{i=1}^{n}\theta_{i}-\sum_{k=1}^{\tilde{Q}}\sum
_{j=1}^{m_{k}}z_{kj}\right)  \label{10}%
\end{equation}
(with $m_{k}=m+1$ for $1\leq k<Q,~m_{k}=m$ for $Q\leq k$) belong to operators
\[
\psi_{Q,\tilde{Q}}(x)~~~~\text{with }\left\{
\begin{array}
[c]{ll}%
\text{charge} & Q\\
\text{spin }\operatorname{mod}1 & Q\tilde{Q}/N\\
\text{statistics factor} & e^{2\pi i\tilde{Q}/N}%
\end{array}
\right.  .
\]
This follows from (v) and (\ref{8}). Also (\ref{7}) holds which means that the
form factor equations (i) -- (v) are satisfied. In particular we have for
\[%
\begin{tabular}
[c]{lll}%
$\tilde{Q}=0$ & the order parameters & $\sigma_{Q}(x)$\\
$Q=0$ & the disorder parameter & $\mu_{\tilde{Q}}(x)$\\
$Q=\tilde{Q}$ & the para-fermi fields & $\psi_{Q}(x)$%
\end{tabular}
\]
with the commutation rules (\ref{cr}). One can show \cite{BK3} that for $N=2$
the formulae (\ref{5}) and (\ref{10}) reproduce the well known results for the
scaling Ising model as mentioned above. Note also that the properties of the
fields for the $Z_{N}$ Ising model are consistent with the results for the
conformal $Z_{N}$ para-fermi field theory \cite{FZ}. For the higher conserved
currents $J_{L}^{\pm}(x)$ see equations (\ref{J1}) and (\ref{J2}) below.

\subparagraph{Examples:}

Up to normalization constants we calculate for general $N$ for the order
parameters
\begin{align*}
\langle\,0\,|\,\sigma_{1}(0)\,|\,p\rangle &  =1\\
\langle\,0\,|\,\sigma_{2}(0)\,|\,p_{1},p_{2}\rangle^{in}  &  =\frac
{F^{Z}(\theta_{12})}{\sinh\frac{1}{2}(\theta_{12}-2\pi i/N)\sinh\frac{1}%
{2}(\theta_{12}+2\pi i/N)}%
\end{align*}
and for $N=3$%
\begin{align*}
\langle\,0\,|\,\sigma_{1}(0)\,|\,p_{1},p_{2},p_{3},p_{4}\rangle^{in}  &
=\left(  \sum e^{\theta_{i}}\sum e^{-\theta_{i}}-1\right) \\
&  \times\prod_{1\leq i<j\leq4}\frac{F^{Z}(\theta_{ij})}{\sinh\frac{1}%
{2}(\theta_{ij}-2\pi i/3)\sinh\frac{1}{2}(\theta_{ij}+2\pi i/3)}%
\end{align*}
which agrees with the results of \cite{K,KiSm}. For disorder parameters we
have
\[
\langle\,0\,|\,\mu_{\tilde{Q}}(0)\,|\,0\rangle=1
\]
and for $N=3$ we obtain
\begin{align*}
\langle\,0\,|\,\mu_{1,2}(0)\,|\,p_{1},p_{2},p_{3}\rangle^{in}  &  =\left(
e^{\pm\theta_{1}}+e^{\pm\theta_{2}}+e^{\pm\theta_{3}}\right)  \,e^{\mp\frac
{1}{3}\sum_{i=1}^{3}\theta_{i}}\\
&  \times\prod_{1\leq i<j\leq3}\frac{F^{Z}(\theta_{ij})}{\sinh\frac{1}%
{2}(\theta_{ij}-2\pi i/3)\sinh\frac{1}{2}(\theta_{ij}+2\pi i/3)}\,.
\end{align*}
For the para-fermi fields we have for example
\begin{align*}
\langle\,0\,|\,\psi_{1}(0)\,|\,p\rangle &  =e^{\frac{1}{N}\theta}\\
\langle\,0\,|\,\psi_{2}(0)\,|\,p_{1},p_{2}\rangle^{in}  &  =\frac{\left(
e^{-\theta_{1}}+e^{-\theta_{2}}\right)  e^{\frac{2}{N}\left(  \theta
_{1}+\theta_{2}\right)  }F^{Z}(\theta_{12})}{\sinh\frac{1}{2}(\theta_{12}-2\pi
i/N)\sinh\frac{1}{2}(\theta_{12}+2\pi i/N)}%
\end{align*}
and for $N=3$%
\begin{align*}
\langle\,0\,|\,\psi_{1}(0)\,|\,p_{1},p_{2},p_{3},p_{4}\rangle^{in}  &
=e^{\frac{2}{3}\sum_{i=1}^{4}\theta_{i}}\sum_{i<j}e^{-\theta_{i}-\theta_{j}}\\
&  \times\prod_{1\leq i<j\leq4}\frac{F^{Z}(\theta_{ij})}{\sinh\frac{1}%
{2}(\theta_{ij}-2\pi i/3)\sinh\frac{1}{2}(\theta_{ij}+2\pi i/3)}\,.
\end{align*}

\paragraph{The affine $A_{N-1}$-Toda model:}

This model possesses only bosonic fields. Therefore we consider constant
p-functions and generalize results of our investigations \cite{BK2} on the
sinh-Gordon model which is the $A_{1}$-Toda model. Again we consider first
form factors for $n$ particles of type $1$. We write the K-function in
(\ref{4}) as a linear combination of the integrals $I_{n\underline{m}}%
^{T}(\underline{\theta},1)$ given by eq.~(\ref{5}) in terms of $\phi^{T}$%
\begin{equation}
K_{n}^{\mathcal{O}}(\underline{\theta})=N_{n}\sum_{m_{1}=0}^{n}\dots
\sum_{m_{N-1}=0}^{n}\prod_{k=1}^{N-1}q_{k}^{m_{k}}I_{n\underline{m}}%
^{T}(\underline{\theta},1) \label{9}%
\end{equation}
for $N-1$ parameters $q_{k}$. In particular $K_{1}(\underline{\theta}%
)=N_{1}\left(  1+\sum_{k=1}^{N-1}q_{k}\right)  $. We propose that this
K-function yields the $n$-particle (charge 1) form factors of the general
exponential of the fields
\[
\mathcal{O}(x)=\,:\!\exp\vec{\gamma}\vec{\varphi}\!:\!(x)\,.
\]
This is motivated as follows: In \cite{BK2} (see also \cite{KW}) we argued
that the form factors of exponentials of bose fields satisfy the momentum
space 'clustering' property for $\operatorname{Re}\theta_{1}\rightarrow
\infty$
\[
\left[  e^{\gamma\varphi}\right]  _{n}(\theta_{1,}\theta_{2,}\dots,\theta
_{n})=\left[  e^{\gamma\varphi}\right]  _{1}(\theta_{1})\,\left[
e^{\gamma\varphi}\right]  _{n-1}(\theta_{2,}\dots,\theta_{n})+O(e^{-\theta
_{1}})\,.
\]
The form factors given by (\ref{4}) and (\ref{9}) show just this asymptotic
behavior because of
\begin{align*}
\lim_{\theta\rightarrow\infty}F^{T}(\theta)  &  =1\\
\lim_{\theta_{1}\rightarrow\infty}I_{n\underline{m}}^{T}(\underline{\theta})
&  =I_{n-1\underline{m}}^{T}(\underline{\theta}^{\prime})+\sum_{k=1}%
^{N-1}I_{n-1\underline{m}_{k}}^{T}(\underline{\theta}^{\prime})
\end{align*}
where $\underline{\theta}^{\prime}=(\theta_{2},\dots,\theta_{n})$ and
$\underline{m}_{k}=m_{1},\dots,m_{k}-1,\dots,m_{N-1}$. The normalization
constants are obtained from this equation and (\ref{l}) as $N_{n}=N_{1}^{n}$
and
\begin{equation}
N_{1}=\left(  \frac{2i}{r}\right)  ^{1/2}\left(  \dfrac{1}{F^{T}(iu)}%
\prod_{k=2}^{N-1}\frac{\left(  \phi^{T}(kiu)\right)  ^{1/2}}{\left(  \phi
^{T}(kiu)F^{T}(kiu)\right)  ^{k}}\prod_{k=1}^{N-1}q_{k}^{-1}\right)  ^{1/N}
\label{n}%
\end{equation}
where $u=\frac{2\pi}{N},\;r=\operatorname*{Res}\limits_{z=iu}\phi^{T}(z)$ and
the intertwiner $\Gamma_{k1}^{k+1}=i\left(  \frac{r}{i\phi^{T}(-kiu)}\right)
^{1/2}$ have been used.

The exponentials of the special linear combination of fields $\vec{\gamma}%
\vec{\varphi}(x)=\gamma\vec{\alpha}_{j}\vec{\varphi}(x)$ are obtained for the
choice
\begin{equation}
q_{k}=\omega^{-k}\omega^{N-\hat{\gamma}\left(  \delta_{kj}-\delta
_{kj+1}\right)  }. \label{q}%
\end{equation}
(For $j=0$ and $N-1$ there are extra factors $\omega^{\pm\hat{\gamma}}$). As
in \cite{BK1,BK2} we obtain the fields $\vec{\alpha}_{j}\vec{\varphi}(x)$ by
expansion in terms of $\gamma$. The quantum version of the field equation
(\ref{fe}) is satisfied for all matrix elements if we take $\hat{\gamma
}=\gamma\frac{B}{2\beta}$ and relate the renormalized and the bare mass by
\[
m^{2}=4\,\alpha\,\sin^{2}\frac{\pi}{N}\frac{2N\sin\frac{\pi}{2N}B\sin\frac
{\pi}{2N}(2-B)}{\pi B\sin\frac{\pi}{N}}%
\]
which agrees with the results of \cite{Lu,ABFKR}. The proof is analog to the
one in \cite{BK2}. The proposal (\ref{q}) is in addition motivated by the fact
that the vacuum 1-particle (of charge $b$) matrix element for the field with
charge $a$ turns out to be proportional to $\delta_{ab}$ \footnote{This
corresponds to the complex representation of the roots $\alpha_{j}^{a}%
=\frac{-1}{\sqrt{N}}2\sin\frac{a\pi}{N}\omega^{-ja}$ and fields with $\left(
\varphi^{a}\right)  ^{\dagger}=\varphi^{N-a}$.}
\[
\langle\,0\,|\,\varphi^{a}(0)\,|\,p\,\rangle_{b}=\frac{-1}{\sqrt{N}2\sin
\frac{a\pi}{N}}\sum_{j=0}^{N-1}\omega^{ja}\langle\,0\,|\,\vec{\alpha}_{j}%
\vec{\varphi}(0)\,|\,p\,\rangle_{b}=\delta_{ab}\sqrt{Z^{a}}\,.
\]
The form factors of the bound states $|\,p\,\rangle_{b}$ are obtained from
(\ref{9}) with (\ref{q}) by applying iteratively (iv) as%
\begin{align*}
\langle\,0\,|:\!\exp\gamma\vec{\alpha}_{j}\vec{\varphi}\!:\!(0)\,|\,p\,\rangle
_{b}  &  =\frac{\beta}{\pi B}\sqrt{NZ^{b}}\chi_{b}(\gamma)\\
\chi_{b}(\gamma\vec{\alpha}_{j})  &  =\prod_{k=1}^{N-1}q_{k}^{-b/N}\sum_{0\leq
k_{1}<\dots<k_{b}<N}q_{k_{1}}\dots q_{k_{b}}\\
&  =-4\omega^{-bj}\frac{\sin\frac{b\pi}{N}}{\sin\frac{\pi}{N}}\sin\left(
\frac{\pi}{N}\frac{B\gamma}{2\beta}\right)  \sin\frac{\pi}{N}\left(
1-\frac{B\gamma}{2\beta}\right)  \,.
\end{align*}
The function $\chi_{b}$ may be written as a character $\chi_{b}(\gamma
\vec{\alpha}_{j})=\operatorname{tr}_{\pi_{b}}\omega^{\left(  \vec{\rho}%
-\frac{B}{2\beta}\gamma\vec{\alpha}_{j}\right)  \vec{H}}$ \cite{Lu}. By
expansion with respect to $\gamma$ we obtain%
\[
\langle\,0\,|\vec{\alpha}_{j}\vec{\varphi}(0)\,|\,p\,\rangle_{b}=-\sqrt
{\frac{Z^{b}}{N}}\omega^{-bj}\,2\sin\frac{\pi b}{N}\,.
\]
The wave function renormalization constants $Z^{a}$ are calculated from (vi)
and (\ref{n}) as
\[
Z^{a}=\frac{\pi B(2-B)}{4\sin\frac{\pi}{2}B}\exp\int_{0}^{\infty}\frac{dt}%
{t}\,\frac{2\sinh\frac{t}{2}\frac{B}{N}\sinh\frac{t}{2}\frac{2-B}{N}}%
{\sinh^{2}t\sinh t\frac{1}{N}}\left(  \sinh^{2}t\frac{a}{N}+\sinh^{2}%
t\frac{\bar{a}}{N}\right)
\]
which satisfies the charge conjugation symmetry $a\leftrightarrow\bar{a}=N-a$
and agree with the results of \cite{DdV,Lu}.

The higher conserved currents $J_{L}^{\pm}(x)$ which are characteristic for
integrable models are obtained by (\ref{4}) and (\ref{5}) and the K- and
p-function%
\begin{align}
K_{n}^{J_{L}^{\pm}}(\underline{\theta})  &  =\sum_{m_{1}=0}^{n}\dots
\sum_{m_{N-1}=0}^{n}\prod_{k=1}^{N-1}\omega^{km_{k}}I_{n\underline{m}}%
^{T}(\underline{\theta},p_{n}^{J_{L}^{\pm}})\label{J1}\\
p_{n\underline{m}}^{J_{L}^{\pm}}(\underline{\theta},\underline{z})  &
=\sum_{i=1}^{n}e^{\pm\theta_{i}}\sum_{k=1}^{N-1}\sum_{j=1}^{m_{k}}e^{Lz_{kj}%
}\,. \label{J2}%
\end{align}
The higher charges $Q_{L}=\int dxJ_{L}^{0}(x)$ satisfy the eigenvalue
equation
\[
\left(  Q_{L}-\sum_{i=1}^{n}e^{L\theta_{i}}\right)  \,|\,p_{1},\ldots
,p_{n}\,\rangle^{in}=0\,.
\]
Since the currents are $Z_{N}$-chargeless the number $n$ of particles of
charge 1 is $0\operatorname{mod}N$. Obviously we get the energy momentum
tensor from $J_{\pm1}^{\pm}(x)$. The higher conserved currents $J_{L}^{\pm
}(x)$ for the $Z_{N}$-Ising model are obtained by same p-function with the
additional factor $\prod_{k=1}^{N-1}\delta_{m_{k}m}$ for $n=Nm$. In a more
detailed version of this letter \cite{BK3}, we will present the proofs and
further results.

\section*{Acknowledgments}

We thank V.A. Fateev, A. Fring, A. Nersesyan, R. Schrader, B. Schroer, J.
Teschner, A. Tsvelik, Al.B. Zamolodchikov and A.B. Zamolodchikov for
discussions. In particular we thank V.A. Fateev for bringing the preprint
\cite{KiSm} to our attention and F.A. Smirnov for sending a copy. H.B. was
supported by DFG, Sonderforschungsbereich 288 `Differentialgeometrie und
Quantenphysik' and partially by the grants INTAS 99-01459 and INTAS 00-561.
This work is also supported by the EU network EUCLID, 'Integrable models and
applications: from strings to condensed matter', HPRN-CT-2002-00325.


\begin{thebibliography}{99}
\bibitem{KS}R.K\"{o}berle and J.A. Swieca, \emph{Phys. Lett.} \textbf{B }86
(1979) 209.

\bibitem{Za}A.B. Zamolodchikov, \emph{Int. Journ. Mod. Phys., } \textbf{A 3}
(1988) 743-750.

\bibitem{Fa}V.A. Fateev, \emph{Int. Journ. Mod. Phys., } \textbf{A 6} (1991) 2109-2132.

\bibitem{FZ}V.A. Fateev and A.B. Zamolodchikov, \emph{Sov. Phys. JETP, }
\textbf{62} (1985) 215.

\bibitem{B-J}R.Z. Bariev, \emph{Phys. Lett.} \textbf{55A} (1976)
456;\newline B. McCoy, C.A. Tracy and T.T. Wu, \emph{Phys. Rev. Lett.}
\textbf{38} (1977) 783; \newline M. Sato, T. Miwa and M. Jimbo, \emph{Proc.
Japan Acad.} \textbf{53A} (1977) 6.

\bibitem{BKW}B. Berg, M. Karowski and P. Weisz, \emph{Phys. Rev.} \textbf{D19}
(1979) 2477.

\bibitem{AFZ}A.E. Arinshtein, V.A. Fateev and A.B. Zamolodchikov, \emph{Phys.
Lett.} \textbf{B 87} (1979) 389.

\bibitem{BCDS}H.W Braden, E. Corrigan, P.E. Dorey and R. Sasaki, \emph{Nucl.
Phys.} \textbf{B338} (1990) 689-746.

\bibitem{K}M. Karowski, \emph{Field Theories in 1+1 Dimensions with Soliton
Behaviour: Form Factors and Green's Functions}, in 'Lecture Notes in Physics
126' (Springer) (1979) p. 344.

\bibitem{KiSm}A.N. Kirilov and F.A. Smirnov, Kiev preprint ITF-88-13P (1988)
(in Russian).

\bibitem{JKOPS}M. Jimbo, H. Konno, S. Odake, Y. Pugai and J. Shiraishi,
\emph{Free field construction for the ABF models in Regime II, } (2000) math.QA/0001071.

\bibitem{DdV}C. Destri and H.J. De Vega, \emph{Nucl. Phys.} \textbf{B358}
(1991) 251-294.

\bibitem{Oo}T. Oota, \emph{Nucl. Phys.} \textbf{B466} [FS] (1996) 361-382.

\bibitem{Lu}S. Lukyanov, Phys. Lett. \textbf{B408} (1997) 182-200.

\bibitem{CF}O. Castro-Alvaredo and A. Fring, \emph{Nucl. Phys.} \textbf{B618}
(2001) 437.

\bibitem{KW}M. Karowski and P. Weisz, \emph{Nucl. Phys.} \textbf{B139} (1978) 445.

\bibitem{Sm}F.A. Smirnov \emph{'Form Factors in Completely Integrable Models
of Quantum Field Theory', Adv. Series in Math. Phys.} \textbf{14}, World
Scientific 1992.

\bibitem{BFKZ}H. Babujian, A. Fring, M. Karowski and A. Zapletal, \emph{Nucl.
Phys.} \textbf{B538} [FS] (1999) 535-586.

\bibitem{BK}H. Babujian and M. Karowski, \emph{Nucl. Phys.} \textbf{B620}
(2002) 407.

\bibitem{Wa}K.M. Watson, \emph{Phys. Rev.} \textbf{95} (1954) 228.

\bibitem{CT}S. Coleman and H.J. Thun, \emph{Commun. Math. Phys. }\textbf{61}
(1978) 31.

\bibitem{BK3}H. Babujian and M. Karowski, in preparation.

\bibitem{BK2}H. Babujian and M. Karowski, \emph{Journ. Phys. A: Math. Gen.
}\textbf{35 }(2002) 9081-9104.

\bibitem{BK1}H.M. Babujian and M. Karowski, \emph{Phys. Lett.} \textbf{B 411}
(1999) 53-57.

\bibitem{ABFKR}C. Ahn, P. Baseilhac, V. Fateev, C. Kim and C. Rim, \emph{Phys.
Lett.}\textbf{ B481 }(2000).

\bibitem{DM}G. Delfino and G. Mussardo, \emph{Nucl. Phys. }\textbf{B455
}(1995) 724-758.

\bibitem{AMV}C. Acerbi, G. Mussardo and A. Valleriani, \emph{Int. J. Mod.
Phys. }\textbf{A11 }(1996) 5327-5364.

\bibitem{FMS}A. Fring, G. Mussardo and P. Simonetti, \emph{Nucl. Phys.
}\textbf{B393 (}1993) 413-441.

\bibitem{KM}A. Koubek and G. Mussardo, \emph{Phys. Lett. }\textbf{B311 (}1993) 193-201.
\end{thebibliography}
\end{document}